\documentstyle[prb,aps,multicol,epsf]{revtex}

\begin{document}

\preprint{LA-UR-96-1114 {\bf /} GISC-1/96 {\bf /} MA/UC3M/12/96}

\draft

\title{Dynamical phenomena in Fibonacci semiconductor superlattices}

\author{Enrique Diez,$^{1,2,\dag}$
Francisco\ Dom\'{\i}nguez-Adame,$^{2,3,4,\ddag}$
Enrique Maci\'{a},$^{2,3,\S}$ and\\
Angel S\'{a}nchez$^{1,2,4,*}$}

\address{$^1$Departamento de Matem\'aticas, 
Escuela Polit\'ecnica Superior,\\
Universidad Carlos III, E-28911 Legan\'{e}s, Madrid, Spain\\
$^2$Grupo Interdisciplinar de Sistemas
Complicados, Escuela Polit\'ecnica Superior,\\
Universidad Carlos III, E-28911 Legan\'{e}s, Madrid, Spain\\
$^3$Departamento de F\'{\i}sica de Materiales,
Facultad de F\'{\i}sicas,\\
Universidad Complutense, E-28040 Madrid, Spain\\
$^4$Theoretical Division and Center for Nonlinear Studies, \\
Los Alamos National Laboratory, Los Alamos, New Mexico 87545}

\date{\today}

\maketitle

\begin{abstract}

We present a detailed study of the dynamics of electronic wavepackets in
Fibonacci semiconductor superlattices, both in flat band conditions and
subject to homogeneous electric fields perpendicular to the layers.
Coherent propagation of electrons is described by means of a scalar
Hamiltonian using the effective-mass approximation.  We have found that
an initial Gaussian wavepacket is filtered selectively when passing
through the superlattice.  This means that only those components of the
wavepacket whose wavenumber belong to allowed subminibands of the
fractal-like energy spectrum can propagate over the entire superlattice.
The Fourier pattern of the transmitted part of the wavepacket presents
clear evidences of fractality reproducing those of the underlying energy
spectrum.  This phenomenon persists even in the presence of unintentional
disorder due to growth imperfections.  Finally, we have demonstrated
that periodic coherent-field induced oscillations (Bloch oscillations), which we
are able to observe in our simulations of periodic superlattices, are
replaced in Fibonacci superlattices by more complex oscillations displaying
quasiperiodic signatures, thus sheding more light onto the very
peculiar nature of the electronic states in these systems.

\end{abstract}


\begin{multicols}{2}

\narrowtext

\section{Introduction}

Since the fabrication of aperiodic semiconductor superlattices (SLs)
arranged according to the Fibonacci \cite{Merlin1} and Thue-Morse
\cite{Merlin2} sequences, there has been a growing interest in their
electronic properties, both from experimental
\cite{Merlin3,Laruelle,Yamaguchi,Toet,Hirose,Munzar} and theoretical
\cite{Laruelle,Macia1,Adame1,Adame2} viewpoints.  One of the most appealing
motivation for these studies is the theoretical prediction that ideal
aperiodic SLs should exhibit a highly-fragmented electronic spectrum
displaying self-similar patterns.
\cite{Kohmoto,Ryu} And, in fact, photoluminescence excitation
spectroscopy at
low temperature reveals the existence of a fragmented density of states
consistent with theoretical predictions. \cite{Yamaguchi}

Other motivation for the study of dynamical phenomena in aperiodic
systems is the following. Electron
states in periodic SLs spread uniformly over the whole SL (Bloch states)
and the energy spectrum is composed by minibands and minigaps.  In the
absence of electric field, these extended states are characterized by a
transmission probability very close to unity.  When a homogeneous
electric field is applied perpendicular to the layer plane, electronic
states become localized (Stark-Wannier states) and the energy spectrum
consists of equally spaced levels (Stark-Wannier ladder).  From the
perspective of quantum evolution, Bloch states lead to Bloch
oscillations (BOs) when the electric is applied. \cite{Bloch,Dignam} The
BO time period of the electronic motion in real as well as in $k$ space
is given by \cite{Dignam}
\begin{equation}
\tau_{\text{B}} = {2\pi \hbar\over eFd},
\label{period}
\end{equation}
where $d$ is the SL constant.  High-quality SLs make it possible to
obtain BO time periods larger than the scattering time for reasonable
values of the applied electric field $F$.  Reports of unambiguous
experimental evidence for BOs in periodic GaAs-Ga$_{1-y}$Al$_y$As SLs have been
recently appeared, \cite{Leo,Feldmann} using an experimental method
previously proposed by von Plessen and Thomas. \cite{Plessen} This
picture is  assumed to be
no longer valid in Fibonacci SLs (FSLs) because in the
thermodynamical
limit electron states are critical instead of extended in flat band
conditions, from a strict mathematical point of view.  However, since
actual FSLs are of finite size, one could expect similar transport
properties (i.\ e., high transmission coefficient) to that shown by
extended electronic states since they spread over the whole FSL,
although we insist that they are not Bloch states.  Hence the question
as to whether
BOs will be observed or not in FSL arises quite naturally.

Therefore the aim of this work is twofold.  In the first place, we
provide a complete characterization of electronic states in FSLs, giving
a detailed description of dynamical phenomena of electronic wavepackets,
which, as far as we know, have not been reported in the literature.  In
this way, we are led to the conclusion that FSLs act as {\em efficient
electronic filters}, an appealing property in order to use them in
actual devices of technological interest. In the second place, we
investigate the possibility to observe BOs in FSLs, suggesting the
convenience of generalizing the concept of periodic BOs to the case
of {\em quasiperiodic} oscillations in order to properly describe the dynamical
behavior of critical states under the action of homogeneous electric
fields.

To this end,
in this paper we address the study of time-dependent effects in FSLs by
solving numerically the effective-mass equation for the
envelope-function, both in the absence of external fields and under
homogeneous electric fields.  To be specific, we consider the problem of
quantum evolution of electronic wavepackets initially localized in space
impinging on the FSL. The wavepacket dynamics will be properly described
by means of the time-dependent transmission probability.  The
transmitted portion of the wavepacket will be characterized by its
Fourier transform, aiming to search for particular signatures arising
from the scattering event.  To get an estimation of the spreading of the
wavepacket as a function of time, we will use the time-dependent inverse
participation ratio (IPR) as well as the mean-square displacement.
Finally, since unintentional imperfections appear during growth in
actual FSLs, we have analyzed a modified version of our model to
investigate the possible existence of a competition between the
long-range quasiperiodic order and the short-range disorder which could
be detected by our time analysis.

\section{Model}

We consider quantum-well based GaAs-Ga$_{1-y}$Al$_y$As SLs with the same
barrier thickness $b$ in the whole sample.  The thickness of each
quantum-well is $\Delta x_n-b\equiv x_n-x_{n-1}-b$, $x_n$ being the
position of the center of the {\em n\/}th barrier and $x$ the growth
direction.  We will focus on electronic states close to the bandgap with
${\bf k}_{\perp}=0$ and neglect nonparabolicity effects hereafter, so
that the Ben Daniel-Duke Hamiltonian suffices to describe those states.
The envelope-functions for electron wavepackets satisfy the following
quantum-evolution equation
\begin{equation}
i\hbar\frac{\partial\Psi(x,t)}{\partial t} = {\cal H}(x) \Psi(x,t).
\label{1}
\end{equation}
The time-independent Hamiltonian ${\cal H}(x)$ is given by
\begin{equation}
{\cal H}(x) = -\,{\hbar^2\over 2m^*}\,{d^2\phantom{x}\over dx^2} +
V_{\text{SL}}(x) - eFx,
\label{2}
\end{equation}
where $V_{\text{SL}}(x)$ is the SL potential under flat bands condition and
$F$ is the electric field.  The height of the barrier for electrons is
given by the conduction-band offset at the interfaces $\Delta E_c$.  We
take the origin of electron energies at the GaAs conduction-band edge.

\begin{figure}
\setlength{\epsfxsize}{7.0cm}
\centerline{\mbox{\epsffile{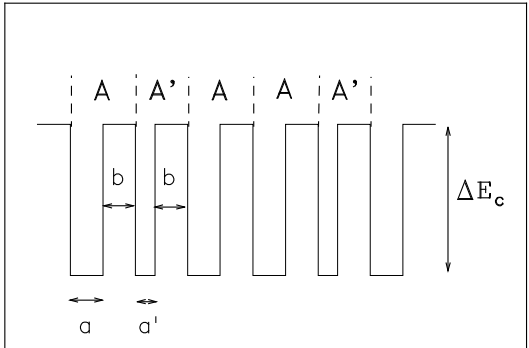}}}
\caption{Schematic diagram of the conduction-band edge of the
GaAs-Ga$_{1-y}$Al$_{y}$As Fibonacci superlattice.}
\label{fig0}
\end{figure}

The particular kind of SLs subject of this work, {}FSLs, can be grown
starting from two basic building blocks $A$ and $A'$ by means of
molecular beam epitaxy. \cite{Merlin1} Here $A$ ($A'$) consists of a
quantum-well of thickness $a$ ($a'$) and a barrier of thickness $b$.
The Fibonacci sequence $S_n$ is generated by appending the $n-2$
sequence to the $n-1$ one, i.e., $S_n=S_nS_{n-1}$.  This construction
algorithm requires initial conditions which are chosen to be $S_0=A$ and
$S_1=A'$.  In this way, finite and self-similar quasiperiodic SLs are
obtained by $n$ successive applications of these rules
leading to the ordering $A\,A'\,A\,A\,A'\,A\ldots$ ,
containing $N=F_n$ barriers 
The
Fibonacci numbers are generated from the recurrence law
$F_n=F_{n-1}+F_{n-2}$, starting with $F_0=F_1=1$.  A few
blocks of the resulting SL potential $V_{\text{SL}}$ are shown in
Fig.~\ref{fig0}.

Unintentional disorder appearing during growth in {\em actual} SLs
depends critically on the growth conditions and it is unknown in most
cases.  Therefore, one is forced to develop a simple model, making
reasonable assumptions on the type of disorder for each particular
sample.  For instance, islands protruding from one semiconductor into
the other cause in-plane disorder and break translational invariance
parallel to the heterojunction.  If the in-plane average size of these
protrusions is much larger than the mean-free-path, then carriers only
{\em see} an ensemble of different layer thicknesses. \cite{Mader} We
model local excess or defect of monolayers by allowing $\Delta x_n$ to
fluctuate uniformly around the nominal values $a+b$ or $a'+b$.  For
definiteness we take $\Delta x_n = a(1+W\epsilon_n)+b$ or $\Delta x_n =
a'(1+W\epsilon_n)+b$, where $W$ is a positive parameter measuring the
maximum fluctuation and $\epsilon_n$'s are distributed according to a
uniform probability distribution $P(\epsilon_n)=1$ if $|\epsilon_n|<1/2$
and zero otherwise.  Note that $\epsilon_n$ is a random variable, even
when the mean values of $\Delta x_n$ follow the Fibonacci sequence.

\section{Numerical Analysis}

We study the quantum dynamics of an initial Gaussian
wavepacket
\begin{equation}
\Psi(x,0) = \left[2 \pi (\Delta X)^2\right]^{-1/4}\,
\exp{\left[\frac{ik_0x-(x-x_0)^2}{4(\Delta X)^2}\right]},
\label{3}
\end{equation}
impinging on the FSL, where the mean kinetic energy is $\langle E
\rangle =\hbar^2k^2_0/2m^*$ and $\Delta X$ measures the width of the
electron wavepacket.  The solution of Eq.~(\ref{1}) is given by
\begin{equation}
\Psi(x,t)=\exp\left(\frac{i}{\hbar}{\cal H}(x)t\right)\Psi(x,0).
\label{4}
\end{equation}
The finite difference representation of the exponential \cite{recipes}
\begin{equation}
\exp\left(\frac{i}{\hbar}{\cal H}(x)t\right)=
\frac{1-{\displaystyle i\over\displaystyle 2\hbar}\,{\cal H}(x)\delta t}
     {1+{\displaystyle i\over\displaystyle 2\hbar}\,{\cal H}(x)\delta t}
+{\cal O}\left[ (\delta t)^3 \right],
\label{finite}
\end{equation}
where $\delta t$ is the time step, brings a powerful and high-accurate
numerical method. In addition, it ensures probability conservation,
\cite{Bouchard2} which has been used at every time step as a first test
of the accuracy of results.  Boundary conditions read $\Psi(\infty,t) =
\Psi(-\infty,t) = 0$ and we have chosen the FSL length sufficiently
large to be sure than the wavepacket never comes close to the boundaries.

Transmission properties of the electronic wavepacket can be successfully
analyzed by means of the time-dependent transmission probability
$P_T(t)$, which is nothing but the probability that at time $t$ the
electron is found to have crossed the whole SL,
\begin{equation}
P_T(t) = \int_{L}^{\infty}dx\,|\Psi(x,t)|^2,
\label{5}
\end{equation}
where $L$ is the length of the system.  In addition, to get a complete
characterization of the motion of the wavepacket, we use the
time-dependent inverse participation ratio, IPR($t$), and the
mean-square displacement, $\sigma(t)$.  The IPR is defined as
\begin{equation}
\mbox{IPR}(t) = \int_{-\infty}^{\infty}dx\,|\Psi(x,t)|^4,
\label{6}
\end{equation}
and it gives an estimation of the spatial extent and the degree of
localization of electronic wavepackets, which can indeed provide very
much information.  Delocalized states are expected to present small IPR
(in the ballistic limit, without applied field, it vanishes as
$t^{-1}$), while localized states have larger IPR. The mean-square
displacement describes how quantum diffusion of wavepackets initially
located in the middle of the FSL takes place.  The mean-square
displacement $\sigma(t)$ is defined as
\begin{equation}
\sigma^2(t) = \int_{-\infty}^{\infty}\> \left( x-\overline{x}\right)^2\,
|\psi(x,t)|^2\,dx.
\label{7}
\end{equation}
with
$$
\overline{x} = \int_{-\infty}^{\infty} \> x \,|\psi(x,t)|^2\,dx.
$$
In the asymptotic regime ($t\to\infty$) one expects $\sigma^2(t) \sim
t^{\gamma}$.  The exponent is $0< \gamma <1$ for localized states,
$\gamma = 1$ for ordinary diffusion, $1< \gamma <2$ for
super-diffusion, and $\gamma = 2$ for ballistic regime.

\section{Results and discussions}

\subsection{Zero field behavior}

As a typical SL we have chosen a GaAs-Ga$_{0.65}$Al$_{0.35}$As
structure, for which the conduction-band offset is $\Delta E_c=0.25\,$eV
and the effective-mass is $m^{*}=0.067m$, $m$ being the free electron
mass.
In our computations we have taken $a=b=32\,$\AA\ and
$a^{\prime}=26\,$\AA.  With our chosen parameters there exists only one
miniband below the barrier in periodic SLs, ranging from $0.102\,$eV up
to $0.177\,$eV.  Thus the miniband width is much larger than the exciton
binding energy, which amounts $\sim 0.01\,$eV in the present SLs.
This is a relevant fact, for it
has been shown that electronic localization (Stark-Wannier
states) is suppressed for low miniband widths. Finally,
to compare with actual SL's, we have considered the
parameter $W$, which governs the imperfection magnitude, ranging from
$0$ up to a maximum of $0.05$.  This value amounts to having protrusions
thicknesses of half a monolayer on average.

\begin{figure}
\setlength{\epsfxsize}{7.0cm}
\centerline{\mbox{\epsffile{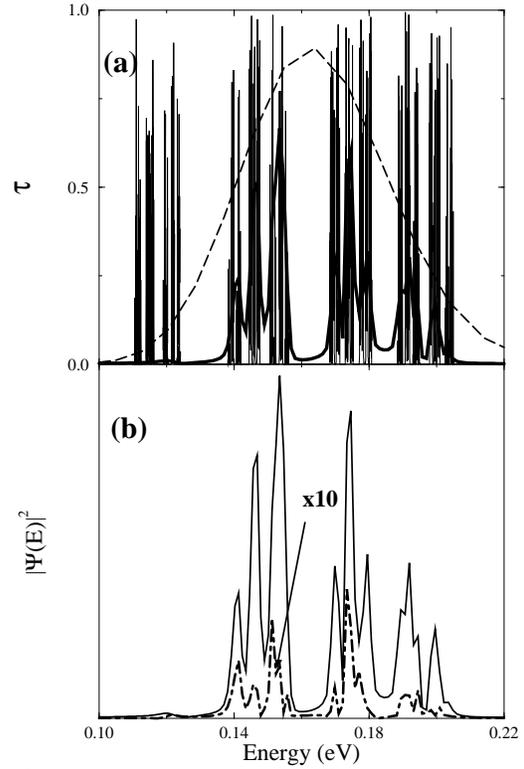}}}
\caption{(a) Transmission coefficient versus incoming energy for a
GaAs-Ga$_{0.65}$Al$_{0.35}$As FSL with $N=144$ wells (thin solid line).
Fourier transforms of the initial wavepacket ($\Delta X=200\,$\AA\ and
$\langle E \rangle = 0.160\,$eV, dashed line) and the transmitted
wavepacket for perfect FSL ($W=0$, thick solid line) at $t=6\,$ps are
also shown in arbitrary units.  (b) Fourier transforms of the trasmitted
wavepackets for a perfect ($W=0$, solid line) and imperfect ($W=0.05$,
dashed line) FSLs at $t=6\,$ps.  Notice in (a) the perfect coincidence
between the filtered wavepacket and the allowed minibands.  Parameters
are $a=b=32\,$\AA\ and $a^{\prime}=26\,$\AA.}
\label{fig1}
\end{figure}

{}Figure~\ref{fig1} collects the results of a typical simulation of a
wavepacket for a FSL. First, in Fig.~\ref{fig1}(a) we show the
transmission coefficient, $\tau$, as a function of energy for a perfect
($W=0$) FSL with $N=144$ wells, computed by means of the transfer-matrix
formalism. \cite{Adame3} The overall structure of the energy spectrum is
characterized by the presence of four main subminibands (notice that
transmission peaks are clustered around energies $\sim 0.120$, $0.155$,
$0.175$, and $0.200\,$eV).  An enlarged view of each cluster of peaks
shows that the fragmentation pattern follows a trifurcation scheme in
which each cluster splits from one to three subclusters upon going to
higher-order generations of FSL's.  This splitting scheme agrees with
that previously reported in FLSs. \cite{Macia1,Macia2} The main question
is to know how this highly-fragmented energy spectrum affects the
quantum evolution of a wavepacket incident on the FSL.

\begin{figure}
\setlength{\epsfxsize}{7.0cm}
\centerline{\mbox{\epsffile{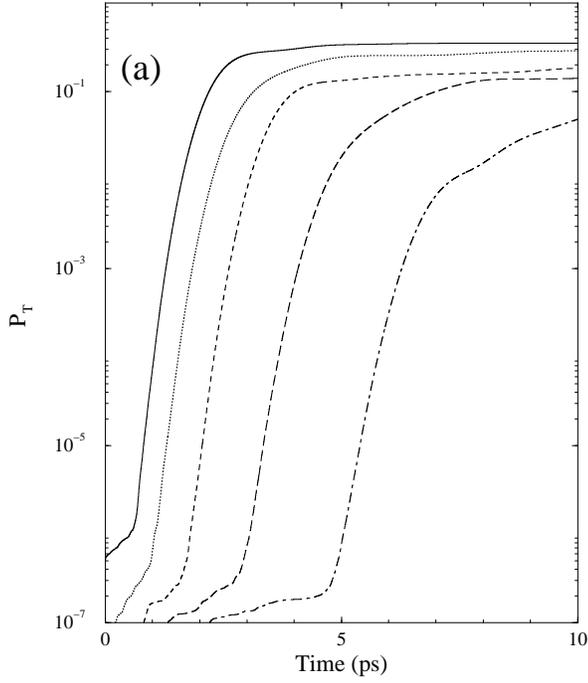}}}
\setlength{\epsfxsize}{7.0cm}
\centerline{\mbox{\epsffile{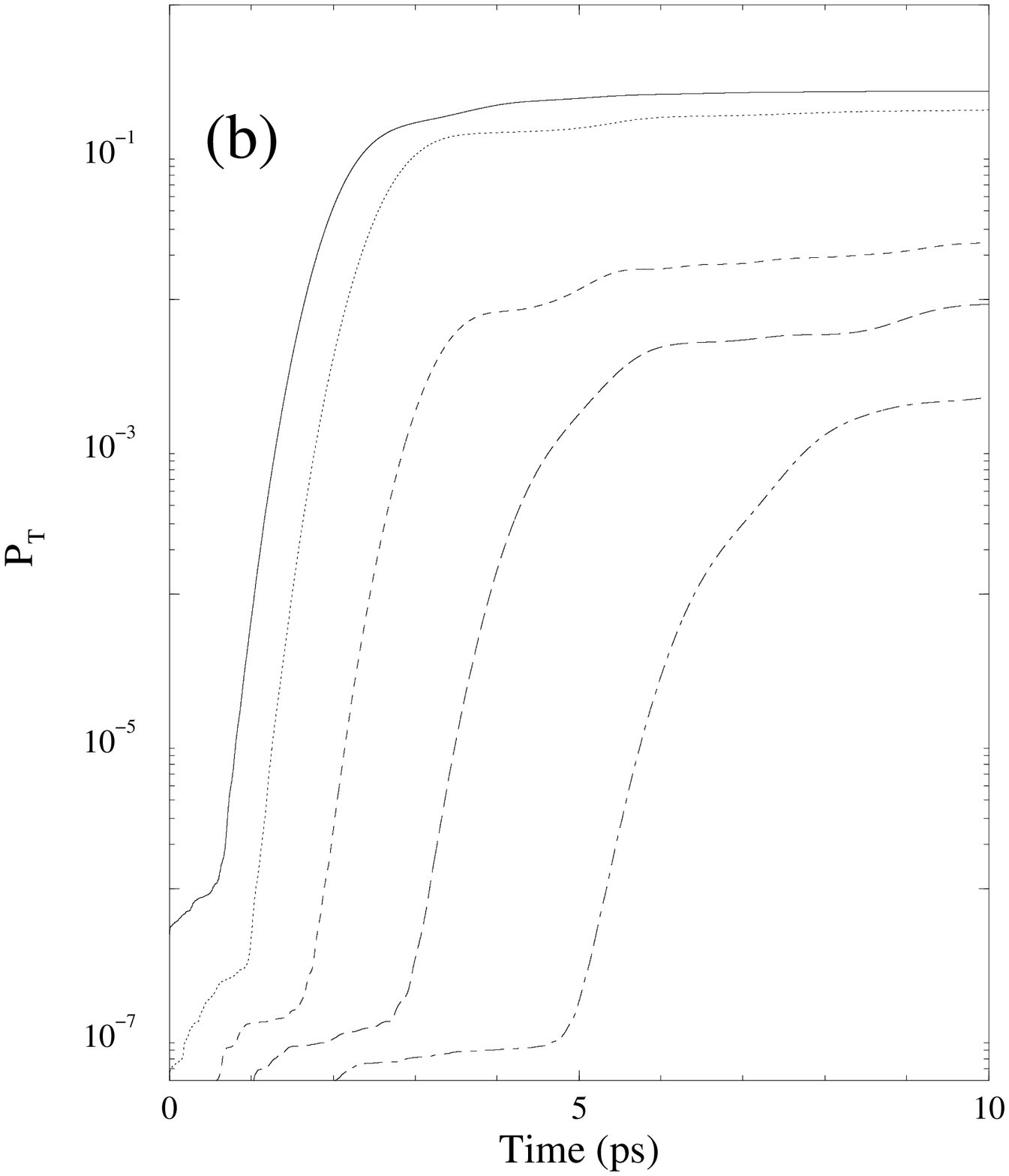}}}
\caption{Transmission probability versus time for (a) perfect $W=0$ and
(b) imperfect $W=0.05$ GaAs-Ga$_{0.65}$Al$_{0.35}$As FSL with various
number of wells: From top to bottom $N=34,55,89,144,233$. Other
parameters are the same as in Fig.~2}
\label{fig2}
\end{figure}

{}Figure~\ref{fig1}(a) also shows the Fourier transform of an initial
Gaussian wavepacket at $t=0$ with its average kinetic energy $\langle E
\rangle = 0.160\,$eV lying in the centermost subminigap and spatial
width $\Delta X = 200\,$\AA.
After the transmitted electron is found to
have crossed the whole FSL [$\sim 6\,$ps, see Fig.~\ref{fig2}(a)], the
Fourier transform of
the electronic wavepacket changes dramatically in perfect FSLs ($W=0$).
Instead of a smooth function, the Fourier transform presents a series of
marked peaks.  Conspicuously, the energy of these peaks coincides with
the higher values of the transmission coefficient, thus indicating that
{\em the FSL acts as an efficient electronic filter}.  Notice that the
Fourier
transform also displays the same splitting pattern than the energy
spectrum, being observable even the third level of hierarchy in the
upper subminiband.  A physical understanding of this behavior is
achieved
if one considers that the initially localized wavepacket can be regarded
as a combination of plane waves in a continuous band.  Since the
dispersion relation (energy versus wavenumber) is self-similar with a
hierarchy of splitted subminibands separated by well-defined minigaps
[see the transmission coefficient shown in Fig.~\ref{fig1}(a)], only
those components whose wavenumber belongs to an allowed subminiband can
propagate over large distances and, consequently, contributing to the
trasmitted part of the wavepacket.  What it is most important for
practical purposes, we have found that unintentional disorder does not
severely affect filtering properties, as shown in Fig.~\ref{fig1}(b) for
$W=0.05$.  Although an overall reduction of the transmitted components
is found, signatures of the above mentioned level splitting are still
clearly observed in the Fourier patter,
particularly at the central energy region around $0.155\,$eV.

\begin{figure}
\setlength{\epsfxsize}{7.0cm}
\centerline{\mbox{\epsffile{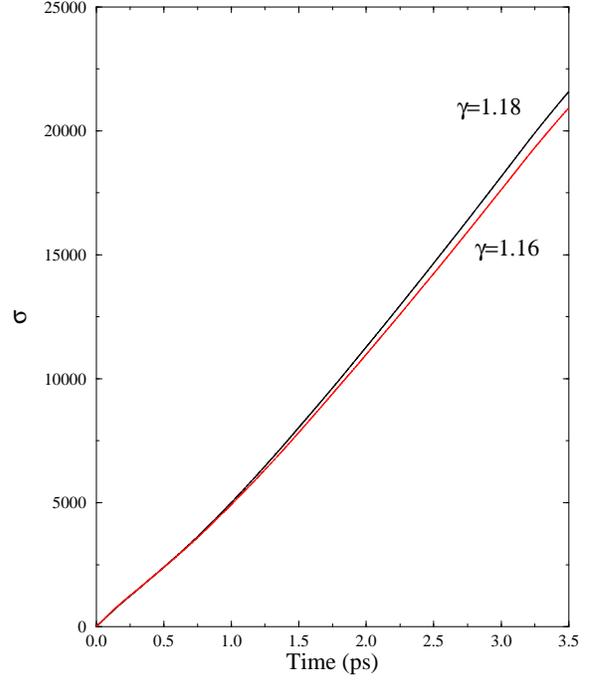}}}
\caption{Mean-square displacement versus time for perfect ($W=0$, solid
line) and imperfect ($W=0.05$, dashed line)
GaAs-Ga$_{0.65}$Al$_{0.35}$As FSL with $N=377$ wells.  Other parameters
are the same as in Fig.~2.  The mean-square displacement grows in time
as a power law $\sim t^{\gamma}$.}
\label{fig3}
\end{figure}

Quantum evolution of electronic wavepackets will depend upon the system
length since the fragmentation of the energy spectrum is higher on
increasing $N$.  Figure~\ref{fig2}(a) shows the time-dependent
transmission probability as a function of time for perfect ($W=0$) FSLs
of various lengths. For comparison, it should be borne in mind that
the transmission probability vanishes in intentionally disordered
SLs for moderately large sizes, thus providing further evidences of
the differences between random and aperiodic systems.
The occurrence of the plateau for larger times
indicates that the transmitted wavepacket has crossed the whole FSL.
Thus, the larger the FSL length, the later the plateau appears, as was
to be expected.  Interestingly, the value of the asymptotic probability
decreases on increasing the FLS length.  This reduction of the
transmission probability means that filtering effects are stronger as
the fragmentation of the energy spectrum is higher.  That is to say, the
equivalent miniband-width, defined as the sum of all the allowed
subminibands, decreases as a power of $N$ as a consequence of the
quasiperiodic topology of the FSL \cite{Adame2} and, therefore, more and
more components of the wavepacket are back reflected.
Figure~\ref{fig2}(b) presents the results for the time-dependent
transmission probability as a function of time for imperfect ($W=0.05$)
FSLs of various lengths.  A comparison with Fig.~\ref{fig2}(a) indicates
a decrease of the transmission probability since short-range disorder
tends to localize particles, according to the Anderson theory.  However,
this effect is almost unnoticeable for relatively short (say $N$
smaller than 89) FSL, in agreement with our previous estimations based
on the study of the equivalent miniband-width. \cite{Macia3} Therefore,
we are led to the conclusion that moderately large fluctuations cannot
destroy filtering effects of actual FSLs and that they can safely be
used for practical purposes.
The study of the mean-square displacement in perfect and
imperfect FSLs provides additional support in this sense.
We have placed the initial Gaussian wavepacket in the
middle of the FSL, with $\Delta X=20\,$\AA\ and $\langle E \rangle =
0.160\,$eV.  In all cases we have observed the super-diffusive regime
$\sigma^2(t) \sim t^{\gamma}$ with $\gamma \simeq 1.2$, as illustrated
in Fig.~\ref{fig3}.
This result shows that the time-dependent properties are quite similar
for both perfect and imperfect FSLs as large as samples containing
$N=377$ wells. It is also interesting to note that the exponent
$\gamma$ obtained in our study, which is based on a realistic
continuous model, is lower than that obtained by considering tight-
binding Hamiltonians. \cite{Brito}

\subsection{Homogeneous field effects}

\begin{figure}
\setlength{\epsfxsize}{7.0cm}
\centerline{\mbox{\epsffile{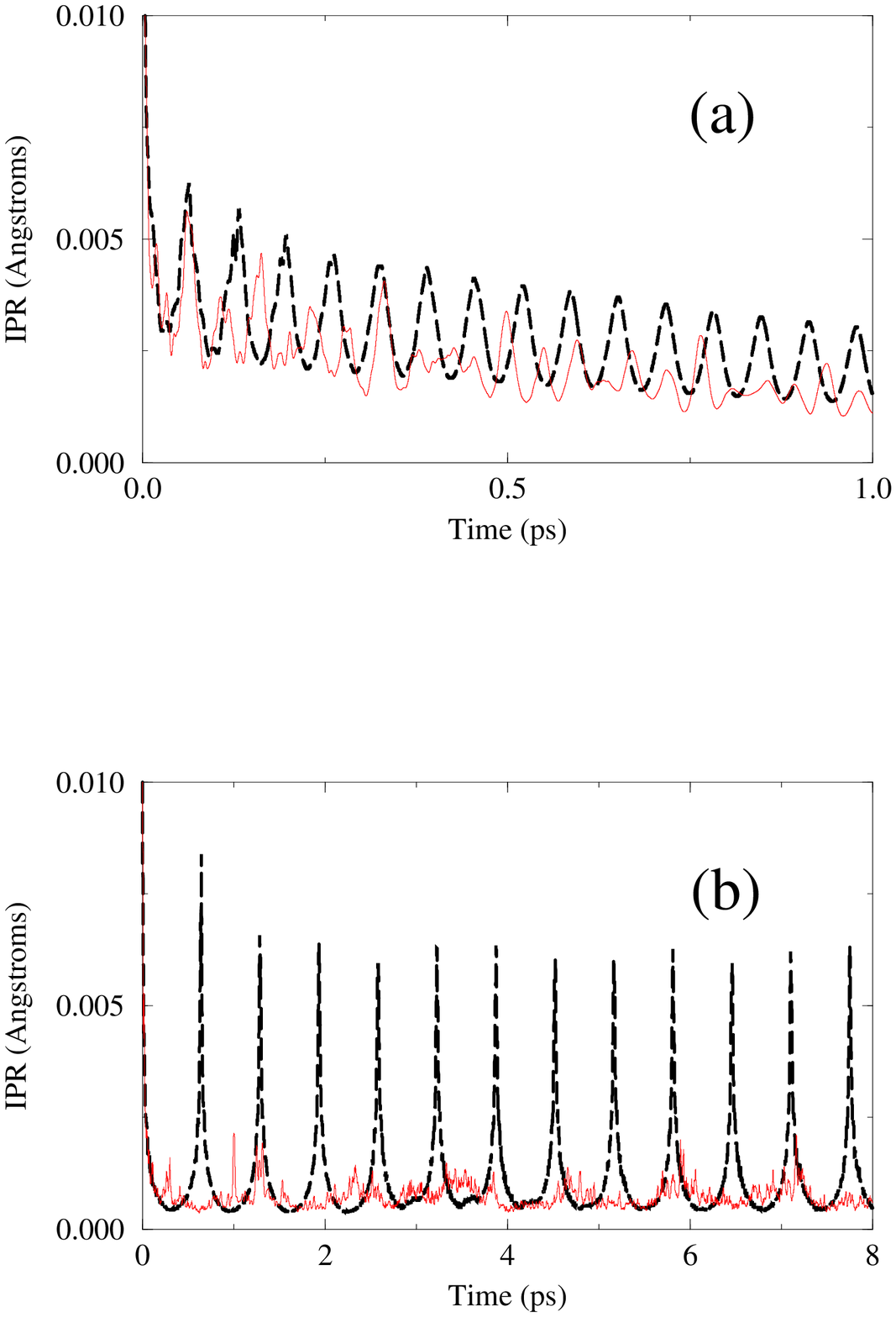}}}
\caption{ Inverse participation ratio as function of time for an initial
Gaussian wavepacket placed in periodic (dashed line) and Fibonacci
(solid line) GaAs-Ga$_{0.65}$Al$_{0.35}$As SLs, subject to an electric
field (a) $F= 100\,$kV/cm and (b) $10\,$kV/cm.  Other parameters are the
same as in Fig.~2.  The occurrence of Bloch oscillations in the periodic
SL is apparent.}
\label{fig4}
\end{figure}

We now comment on our results when a homogeneous
electric field is applied perpendicular to the layers.  One expects that
BOs are to be observed in periodic SLs in the limit of high electric
fields.  The localization length of Stark-Wannier states, the static
counterpart of the BOs, is of the order of $\Delta E/eF$, $\Delta E$
being the width of the allowed miniband, so that we can roughly estimate
that the condition $\Delta E/eF \sim a+b$ establish the high field
regime.  In our SL this field amounts $F\simeq 100\,$kV/cm.
Figure~\ref{fig4}(a) presents the results for the IPR when the initial
Gaussian wavepacket with $\Delta X=20\,$\AA\ and $\langle E \rangle =
0.160\,$eV is located in the middle of the periodic SL, the electric
field being $F=100\,$kV/cm.  The IPR displays periodic oscillations with
marked peaks at times $t_k=k\tau_{\text B}$, where $k$ is any arbitrary,
nonnegative integer and $\tau_{\text B} = 0.065\,$ps.  Thus the IPR is
bounded below, indicating that the wavepacket is localized (dynamical
localization) but its spatial extent varies periodically in time.
Notice that the value of the oscillation period is in excellent
agreement with the theoretical prediction $\tau_{\text{B}} = 2\pi
\hbar/(eFd)$ given in (\ref{period}). It is also worth to mention that
the initial state $\Psi(x,0)$ is not completely restored after each
oscillation, as it should be if interband tunneling were negligible.
\cite{Plessen} Thus we are led to the conclusion that intraband
tunneling plays a role in our periodic SL. 

Results corresponding to a FSL with the same initial condition as before
are also shown in Fig.~\ref{fig4}(a).  First of all, we observe that at
short times we can detect two oscillations coinciding with the
positions of the first two BO peaks, but at larger times {\em periodic}
BOs are completely absent in FSLs.  The absence of Bloch oscillations in
FSLs simply reflects the fact that their extended states are no longer
Bloch states.
Bloch states are characterized by a periodic pattern, but this is not the
case in the FSL, where critical states spreading over the whole system
show quasiperiodic patterns. \cite{Macia4} In this sense, signatures of
of a quasiperiodic oscillation pattern in the IPR corresponding to the
FSL can be seen from Fig.~\ref{fig4}(a) in the interval $0.5\leq t\leq
1\,$ps. To this end, we introduce the ratio $p/q$, where $q$ is the
number of oscillations of the FSL IPR comprised in a given number $p$
of periodic BOs. In this way we obtain the following sequence
$p/q=2/3,3/5,2/3,5/8,2/3,7/11,2/3 \ldots$
which converges to $2/3$, an approximant of the inverse golden mean
$\tau_{G}\equiv (\sqrt{5}-1)/2$.

This facts can be described by means of the following scenario,
based on
our previous discussion on the localization degree of
electronic states under an applied electric field.  For such high
electric fields the spatial extent of the electronic states is of the
order of the SL constant and, consequently, the electron cannot {\em
see} the long-range quasiperiodic potential.  In fact, differences
between periodic and Fibonacci SLs at short times are actually very
small [see Fig.~\ref{fig4}(a)].  As soon as some components escapes from
the localization region (for instance, those states above the barrier
has a high transmission coefficient) the effect of quasiperiodicity takes
place.  To get further confirmation of this assertion we have also
studied the case of lower electric fields, as shown in
Fig.~\ref{fig4}(b) for $F= 10\,$kV/cm, when the localization length is
about ten SL periods. It becomes apparent that no signatures of BOs can
be detected even at short times. Moreover, Fig.~\ref{fig4}(b) also
displays the result for the periodic SL and the same electric field.
Notice that the initial states is completely restored after each BO, the
suggesting that in this regime interband tunneling is actually
negligible.

\section{Conclusion}

In this paper we have studied quantum dynamics of wavepackets in
Fibonacci SLs. After solving numerically the time-dependent
effective-mass equation arising from the Ben Daniel-Duke Hamiltonian, we
have found that an initial Gaussian wavepacket undergoes complicate
scattering events at the FSL. In particular, the Fourier spectrum of the
transmitted part of the wavepacket is no longer smooth but presents many
marked peaks.  These peaks correspond to the allowed subminibands of the
energy spectrum, thus being a clear indication that the FSL acts as a
fine electronic filter.  Transmission properties under zero bias have
also been successfully characterized by means of the transmission
probability given by (\ref{5}).  From these results we have learned that
unintentional disorder arising during growth have no noticeable effects
on the filtering capabilities of short FSLs, because they only cause an
overall decrease of the transmitted amplitude.  On the other hand, the
spatial degree of localization of wavepackets driven by an electric
field has been properly described by means of the time-dependent IPR: As
a check, in periodic SLs we have obtained the dynamical localization
fields as well as BO's.  On the contrary, quantum dynamics in FSLs also
exhibits dynamical localization although it turns out to be much more
intricate.  In particular, no evidence of periodic Bloch oscillations
was observed. Instead, the very concept of BO should be revisited in
order to include the possibility of {\em quasiperiodic} pattern
oscillations emerging from the
quasiperiodic nature of the system.  This result puts another piece of
evidence about the very peculiar nature of electronic states in
Fibonacci systems. \cite{Macia4}

As a concluding remark, we want to stress that with this work we have
sufficiently demonstrated the existence of distinct physical observable
consequences of singular energy spectra like that of the FSL. For
instance, the fragmentation of the SL minibands could serve as the basis
for electronic filters, in a similar fashion to metallic Fibonacci
multilayers acting as selective filters of soft X-ray radiation.
\cite{Adame4} It is quite clear that the output of such a device is but
an image of its fractal spectrum, which, in turn, is intimately
connected with
the quasiperiodic nature of the SL and hence with the information it
contains.  Therefore, aside from the possibility of building the
specific filter-like devices we already mentioned, designed with this or
other quasiperiodic sequence according to the desired application, it
might as well be that this kind of systems can be used in transmitting
or processing information.  This and other relations already established
between quasiperiodic nanoelectronic devices and information
theory\cite{Macia2} pave the way to a very exciting and promising line
of cross-disciplinary research.

\acknowledgments

F.\ D.-A.\ and A.\ S.\ are thankful to Alan R.\ Bishop for the warm 
hospitality enjoyed during their stay at Los Alamos.
Work at Madrid and Legan\'es 
has been supported by CICYT (Spain) under project MAT95-0325.
Work at Los Alamos is performed under the auspices of the U.S. D.o.E.



\end{multicols}

\end{document}